\pdfoutput=1 
\documentclass{JINST} 
\usepackage{graphicx} 
\usepackage{cite}
\usepackage{url}

\title{Performance of Optically Readout GEM-based TPC
with a $^{55}$Fe source}

\author{
{I.~Abritta~Costa}$^{a}$,
{E.~Baracchini}$^{b}$,
{F.~Bellini}$^{a,c}$,
{L.~Benussi}$^{d}$,
{S.~Bianco}$^{d}$,
{G.~Cavoto}$^{a,c}$,
{E.~Di~Marco}$^{a}$,
{G.~Maccarrone}$^{d}$,
{M.~Marafini}$^{e}$,
{G.~Mazzitelli}$^{d}$,
{A.~Messina}$^{a,c}$,
{D.~Piccolo}$^{d}$,
{D.~Pinci}$^{a}$,
{F.~Renga}$^{a}$,
{F.~Rosatelli}$^{d}$ and
{S.~Tomassini}$^{d}$
\\
\llap{$^a$}{Istituto~Nazionale~di~Fisica~Nucleare\\  
 Sezione di Roma, I-00185, Italy\\}
\llap{$^b$}{Gran~Sasso~Science~Institute~L'Aquila, I-67100, Italy \\}
\llap{$^c$}{Dipartimento di Fisica\\  
 Sapienza Universit\`a di Roma, I-00185, Italy\\} 
\llap{$^d$}{Istituto Nazionale di Fisica Nucleare \\  
 Laboratori Nazionali di Frascati, I-00040, Italy\\}
\llap{$^e$}{Museo Storico della Fisica e Centro Studi e Ricerche "Enrico Fermi" \\ 
Piazza del Viminale 1, Roma, I-00184, Italy\\}

E-mail: \email{davide.pinci@roma1.infn.it}}

\date{January 2019}

\abstract{
Optical readout of large Time Projection Chambers (TPCs) with multiple Gas Electron Multipliers (GEMs) amplification stages has shown to provide very interesting performances for high energy particle tracking. Proposed applications for low-energy and rare event studies, such as Dark Matter search, ask for demanding performance in the keV energy range.
The performance of such a readout was studied in details as a function of the electric field configuration and GEM gain by using a $^{55}$Fe source within a 7 litre sensitive volume detector developed as a part of the R\&D for the CYGNUS
project.
Results reported in this paper show that the low noise level of the sensor allows to operate with a 2~keV threshold while keeping a rate of fake-events lesser than 10 per year.

In this configuration, a detection efficiency well above 95\% along with an energy resolution ($\sigma$) of 18\% 
is obtained for the 5.9 keV photons 
demonstrating the very promising capabilities of this technique.}

\begin{document}

\section*{Introduction}

High-resolution tracking for low energy particles had a remarkable development in recent years and will give a crucial contribution in different fields, from medical application to the searches of Dark Matter (DM) massive particles.
A very promising technique involves the optical reading of the
light produced by the de-excitation of gas molecules during the processes of electron multiplication \cite{bib:ref1,bib:ref2,bib:ref3,bib:ref4,bib:nim_orange1,bib:jinst_orange2}. 

This approach has become feasible thanks
to the great progresses achieved in last years in both the performance of Micro Pattern Gas Detectors and the evolution of the CMOS technology which led to the production of sensors with high sensitivity and granularity combined with a very low noise level.

Moreover, the high-resolution tracking capability provided by the optical readout offers the possibility of reconstructing the direction of arrival of the tracks. For application as DM search, this information is very valuable since it provides the possibility of signal identification through topology and direction, very precious to reject background due to internal and external radioactivity \cite{bib:1, bib:2, bib:3, bib:4, bib:ref5}.

The presented R\&D is part of the CYGNUS proto-collaboration effort \cite{bib:0},
aiming at realising a nuclear recoil observatory at the ton scale with
directional sensitivity.

To this aim, the response of a 7 litre sensitive volume detector to 5.9 keV photons produced by a $^{55}$Fe source was studied and the obtained results are reported in this paper.

\section{Experimental setup}
\subsection{LEMON detector}
\begin{figure}[htbp]
\centering
\includegraphics[width=.85\textwidth]{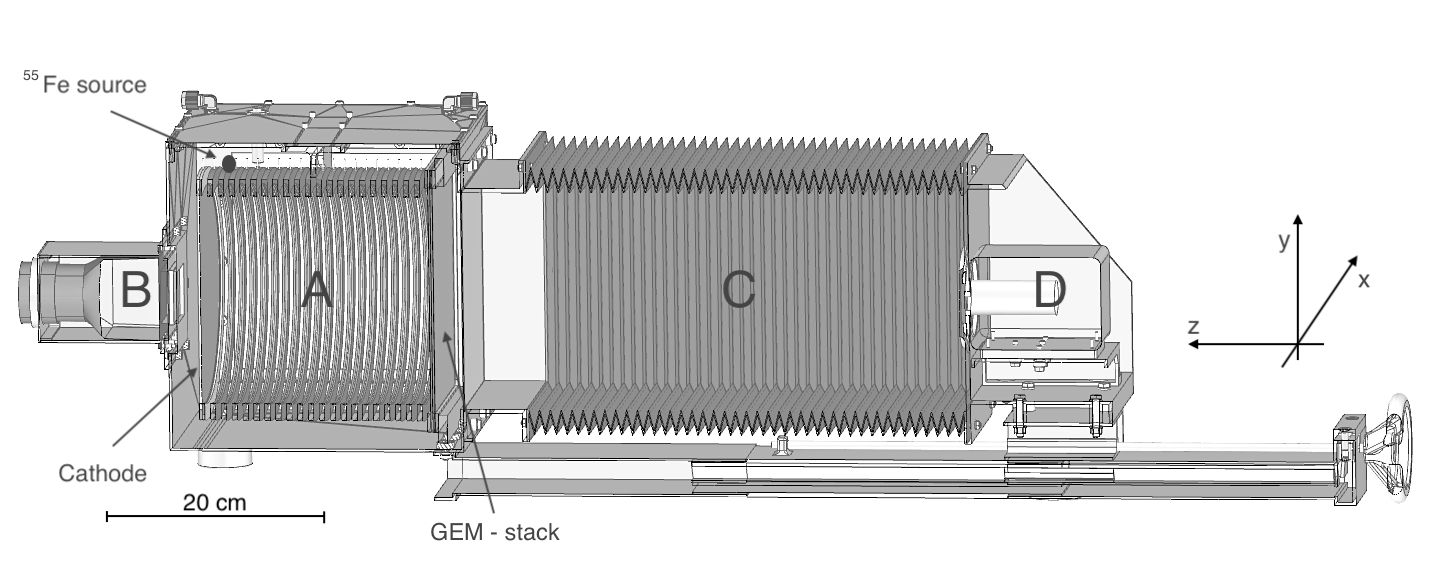}
\caption{Drawing of the experimental setup.
In particular, the elliptical field cage close on one side by the triple-GEM structure and on the other 
side by the semitransparent cathode (A), 
the PMT (B), 
the adaptable bellow (C) and the CMOS camera with its lens (D) 
are visible.}
\label{fig:lemon}
\end{figure}

All measurements described in this work were carried out on the Large Elliptical MOdule (LEMON, Fig. \ref{fig:lemon})
which is described in details in refs. \cite{bib:eps, bib:ieee17, bib:ieee18}.

The main elements of LEMON are:
\begin{itemize}
    \item A sensitive volume (A) filled with 7 litre of He/CF$_4$ 60/40 gas mixture at atmospheric pressure, surrounded by a field cage (FC) composed of 20 elliptic silver plated wire rings with axes of 24 cm (along {\it x}) and 20 cm (along {\it y}) and a depth of 20 cm (along {\it z});
    \item The sensitive volume is closed on one side by a semi-transparent cathode made of a thin wire mesh and on the other side by a structure of three 20$\times$24 cm$^2$, 50~$\mu$m thick GEMs;
    \item The whole structure is contained in a gas-tight box with two transparent windows on the cathode and the GEM sides;
    \item On the cathode side, beyond the window, a fast photo-sensor PMT\footnote{Photonics XP3392, 5 ns rise-time, 76mm square-window} (B) is placed to readout all the light produced by the GEMs;
    \item On the other side, downstream to an adjustable bellow (C), an ORCA Flash 4.0 camera\footnote{For more details visit www.hamamatsu.com}, based on a 1.33~$\times$~1.33~cm$^2$ scientific CMOS sensor (subdivided in 2048~$\times$~2048 pixels with an active area of 6.5~$\times$~6.5~$\mu$m$^2$ each) and equipped with a Schneider lens\footnote{25 mm focal length, 0.95 aperture} (D) is placed at a distance of 52.5~cm (i.e. 21 Focal Length, FL) for the acquisition of the light produced in the GEM holes. In this configuration, the sensor faces a surface of $26~\times~26$~cm$^2$ and therefore each pixel at an area of $130 \times 130~\mu$m$^2$. The geometrical acceptance $\Omega$ therefore results to be $1.6 \times 10^{-4}$ \cite{bib:ieee_orange}.
\end{itemize}

A $^{55}$Fe source, with an activity of about 100 Bq,
was placed between two FC rings, 18 cm far from the GEM as shown in Fig. \ref{fig:lemon}.
Because of the short distance between the plastic rings supporting the FC wires and their width along the {\it x} and {\it y} directions, these acted as a collimator for the photons emitted by the source, so that the effective distance from the GEMs of their interactions with the gas molecules  was estimated to be 18~$\pm$~2 cm. Electrons produced within the sensitive volume are drifted by the electric field (E$_{\rm d}$) present within the FC, toward the GEMs where the multiplication process takes place. Typical operating conditions of the detector are: E$_{\rm d}$~=~600~V/cm, an electric field in the GEM produced by V$_{\rm GEM}$~=~460V for each GEM plane and a transfer field E$_{\rm t}$~=~2~kV/cm. The maximum value of (E$_{\rm d}$) is limited by the maximum voltage provided by the HV generator (15 kV) used for the measurements reported in this paper.

\subsection{Data Acquisition and Analysis}
\label{sec:daq}

\begin{figure}[htbp]
\centering
\includegraphics[width=.45\textwidth]{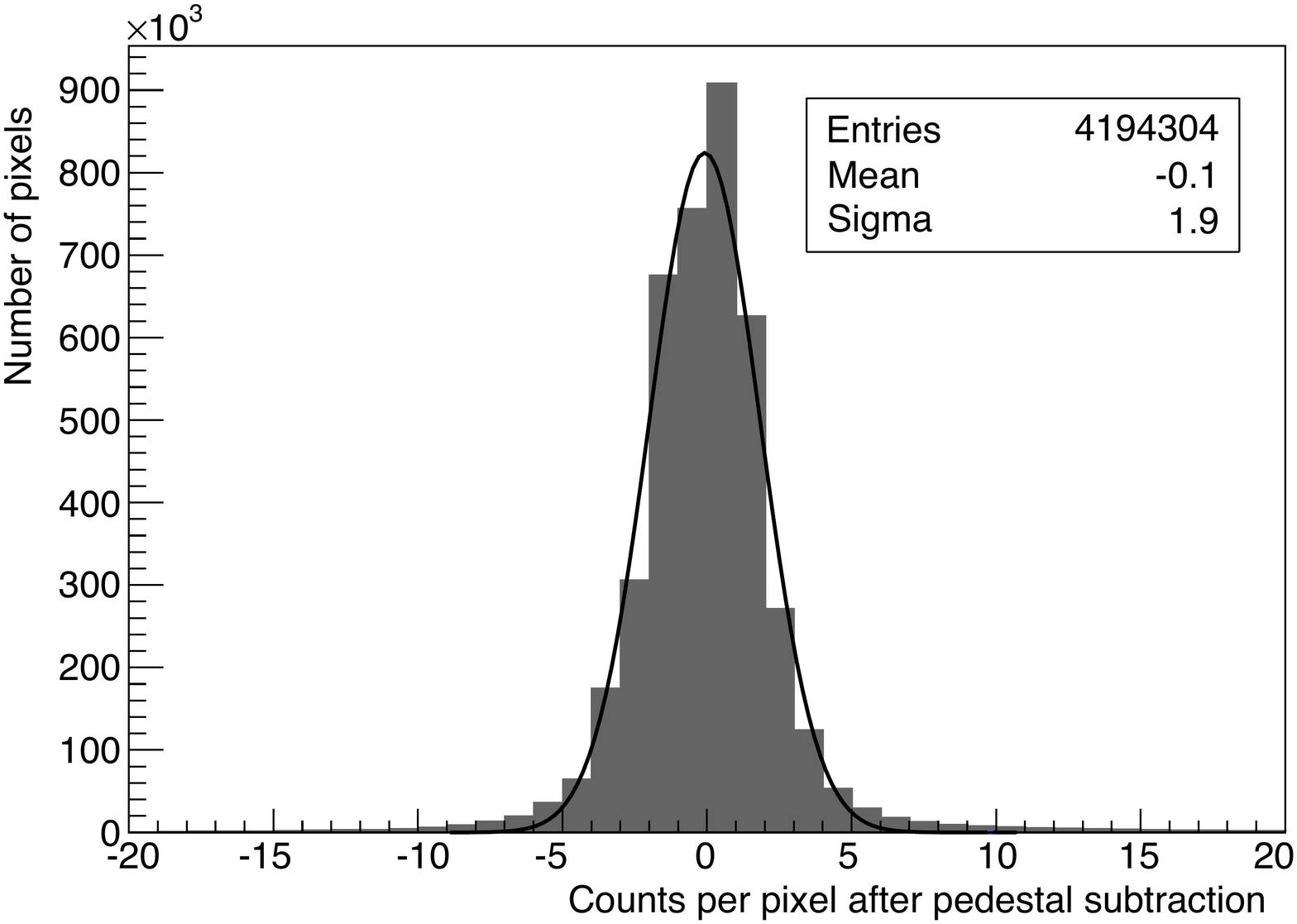}
\caption{Distribution of counts of a blank image once 
the pedestal values were subtracted.}
\label{fig:pedestal}
\end{figure}

Data presented in this paper were acquired by the ORCA sensor 
in free running mode, without any trigger. The light produced during the multiplication processes in the GEM were recorded with an exposure of 100 ms and all data were saved without any selection.
In each configuration, runs of 100 images each were recorded.
The analysis algorithm is based on two steps:
\begin{enumerate}
    \item {\it Pedestal subtraction}. A {\it blind run} of 100 images was acquired with the sensor in total dark. For each pixel, the pedestal is evaluated as the average number of counts recorded in this run and is subtracted to the counts collected in all 
    recorded images. Figure~\ref{fig:pedestal} shows an example of the distribution of the counts in a blank image after the pixel-by-pixel pedestal subtraction. A sensor noise lower than 2 photons per pixel was measured. 
    \item {\it Clustering}: a very simple nearest neighbor-cluster (NNC) clustering algorithm was developed. A lower resolution version of each image was created with {\it macro-pixels} made by matrices of 4$\times$4 pixels.
    The average count over the 16 pixels, after subtracting their pedestal values, is assigned to each macro-pixel.
    A cluster is reconstructed by at least two neighbouring macro-pixels having more than 2 counts (i.e. 2 photons \cite{bib:jinst_orange1}) each. 
\end{enumerate}
For the analysis, three parameters of the reconstructed clusters are studied: total light, number of pixels with more than 2 photons, and position.

\begin{figure}[htbp]
\centering
\includegraphics[width=.45\textwidth]{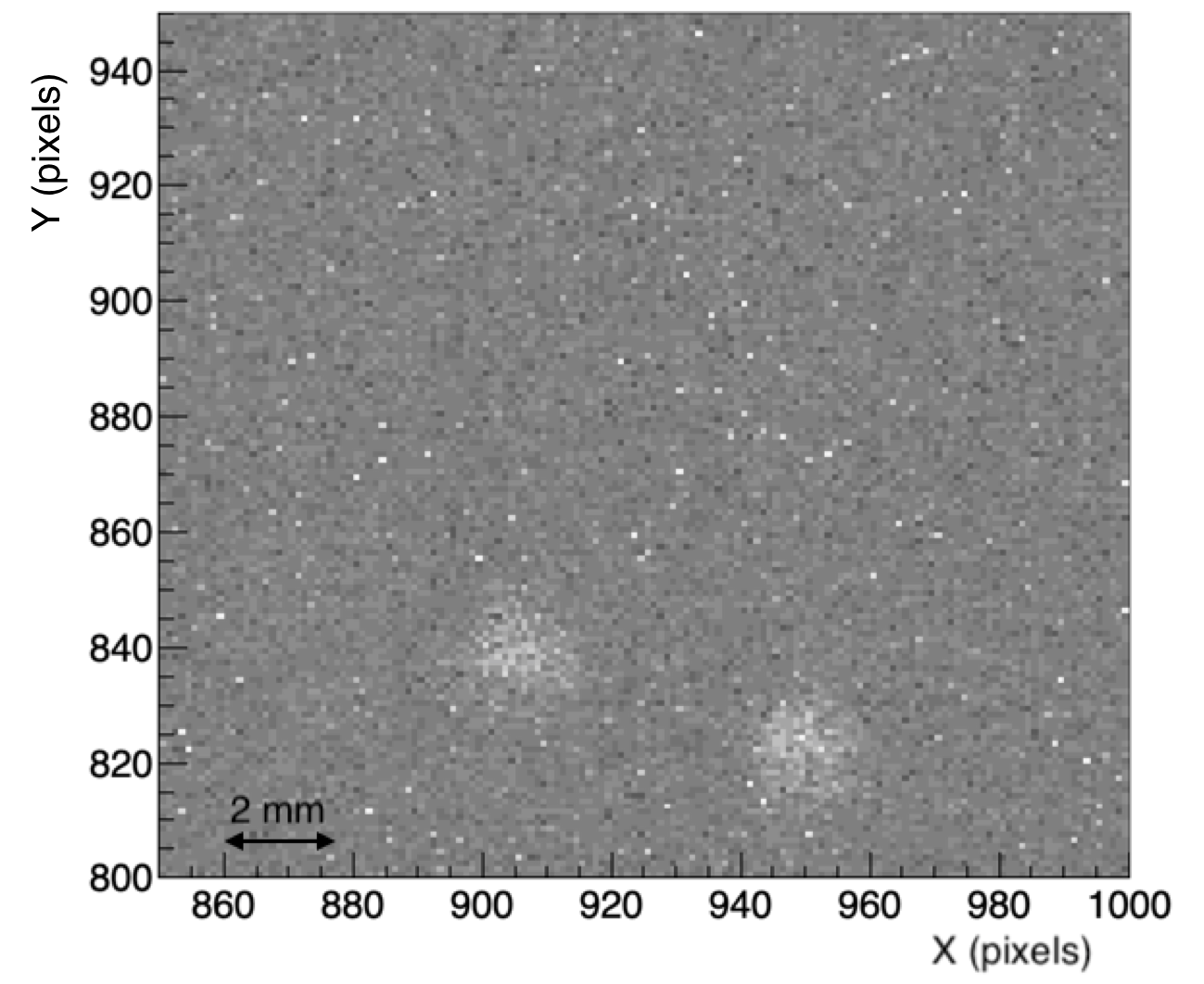}
\caption{Example of two clusters due to X-ray interaction in gas.}
\label{fig:spot}
\end{figure}

Figure~\ref{fig:spot} shows an example of an image of two light 
spots due the interaction of the $^{55}$Fe photons in the gas.

\begin{figure}[htbp]
\centering
\includegraphics[width=.85\textwidth]{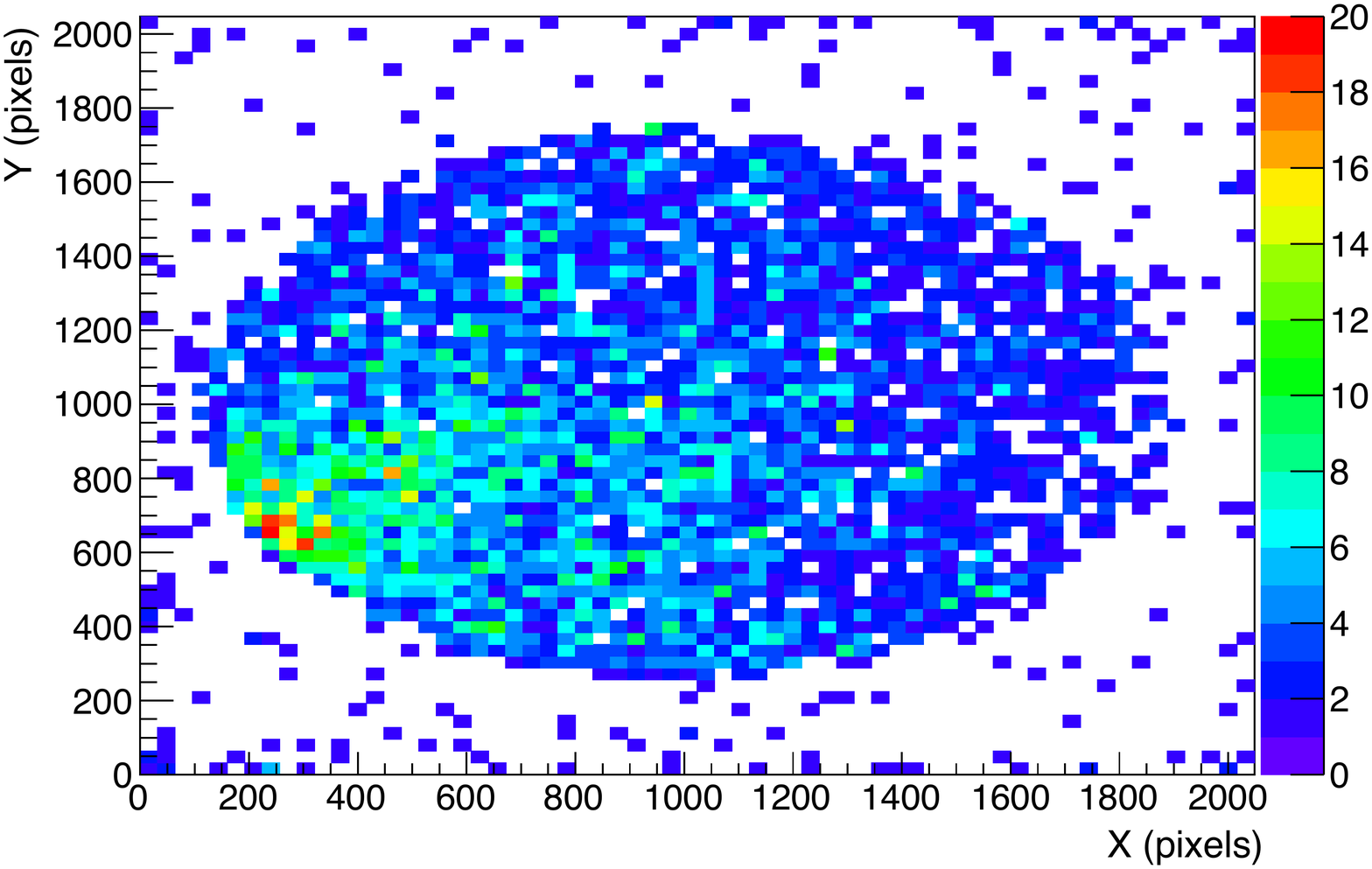}
\caption{Map of the positions of the reconstructed clusters for a run with the $^{55}$Fe source within the detector.}
\label{fig:map_fe}
\end{figure}

The positions of the reconstructed clusters is defined as the average of the positions
of the illuminated pixels in the cluster, weighted for the collected light. An example of the position
distribution in a run is shown in Fig.~\ref{fig:map_fe}. 
Several important features can be recognised:
\begin{itemize}
    \item the position and the shape of the FC is clearly visible. Within it, there is a region at the bottom left with a large number of reconstructed clusters in proximity of the source (signals);
    \item a diffuse background is present, uniformly distributed in the sensitive volume,
    very likely due to natural radioactivity and cosmic rays;
    \item only few clusters are reconstructed, outside the FC, mainly due to the electronic noise of the sensor.
\end{itemize}

\section{Readout noise characterization}
\label{sec:bkg}

\subsection{Sensor electronic noise}
The CMOS sensor used for the measurements has two main sources of noise\footnote{
Consult \url{https://www.hamamatsu.com/resources/pdf/sys/SCAS0134E_C13440-20CU_tec.pdf}}: 
\begin{itemize}
    \item a dark current of about 0.06 electrons per second per pixel;
    \item a readout noise of about 1.4 electrons rms (in our set-up it was found to be slightly
    larger (see Sect. \ref{sec:daq}) probably due to an effect of {\it ageing} of the sensor built more than 5 years ago);
\end{itemize}

\begin{figure}[htbp]
\centering
\includegraphics[width=.45\textwidth]{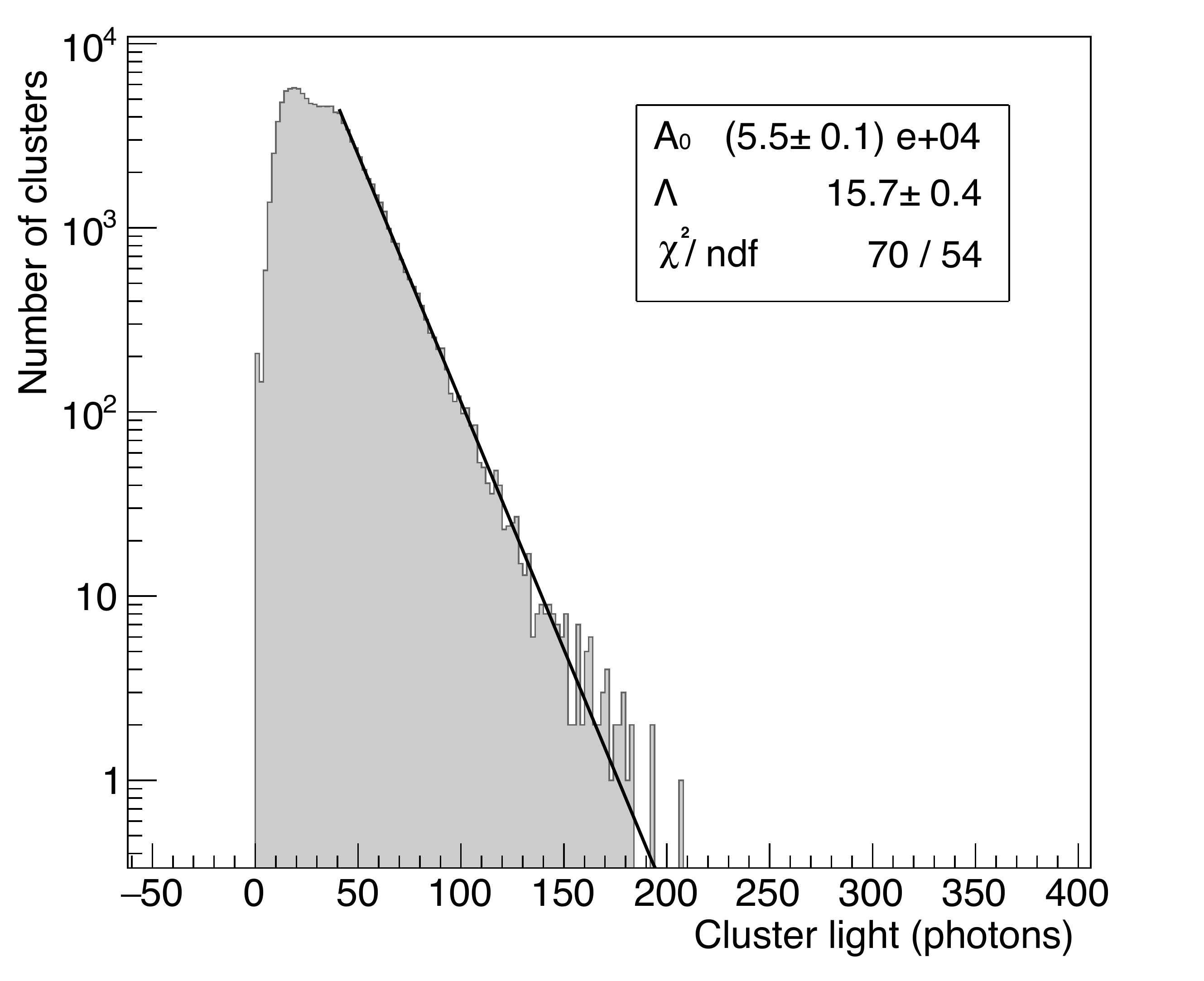}
\caption{Distribution of the light in clusters reconstructed in a run with blind sensor.}
\label{fig:hq_ghost}
\end{figure}

The sensor electronic noise represents a possible
unavoidable instrumental background and it can generate {\it ghost-clusters}.
The distribution of the light in each {\it ghost-cluster} found in the {\it blind run} 
is shown in Fig.~\ref{fig:hq_ghost}. 
It shows a baseline component due to positively-definite counts of photons and to the minimal number of two macro-pixels requested to form a cluster. In order to define an operative threshold that allows to suppress fake signals due to sensor noise, the tail of this distribution is studied. In particular, it is fitted with an exponential function:
\begin{equation}
    p(L)=A_0~e^{-\frac{L}{\Lambda}}
\label{eq:exp}
\end{equation}
where $L$ is the number of photons collected in the cluster. 

From the fitted parameters, $A_0~=~(5.5~\pm~0.1)~\times~10^4$ and $\Lambda=15.7~\pm~0.4$~photons, and by taking into account that a run lasts 10 s, it is possible to extrapolate the probability of having a {\it ghost-clusters} with an amount of light larger than a given threshold. Results for three threshold values are shown in Table~\ref{tab:thr_1}.

\begin{table}[htbp]
\centering
\caption{Number of {\it ghost-clusters} found per sensor per second containing a total amount of light over a certain threshold as extrapolated from the fit in Fig.\ref{fig:hq_ghost}.}
\vspace{0.2 cm}
\begin{tabular}{ |c|c| } 
 \hline
 Threshold (photons) & {\it ghost-clusters}/second \\ 
 \hline
 \hline
 200 & 2$\times$10$^{-2}$  \\ 
 300 & 1$\times$10$^{-4}$  \\ 
 400 & 3$\times$10$^{-7}$  \\ 
 \hline
\end{tabular}
\label{tab:thr_1}
\end{table}

\begin{figure}[htbp]
\centering
\includegraphics[width=.45\textwidth, angle=270]{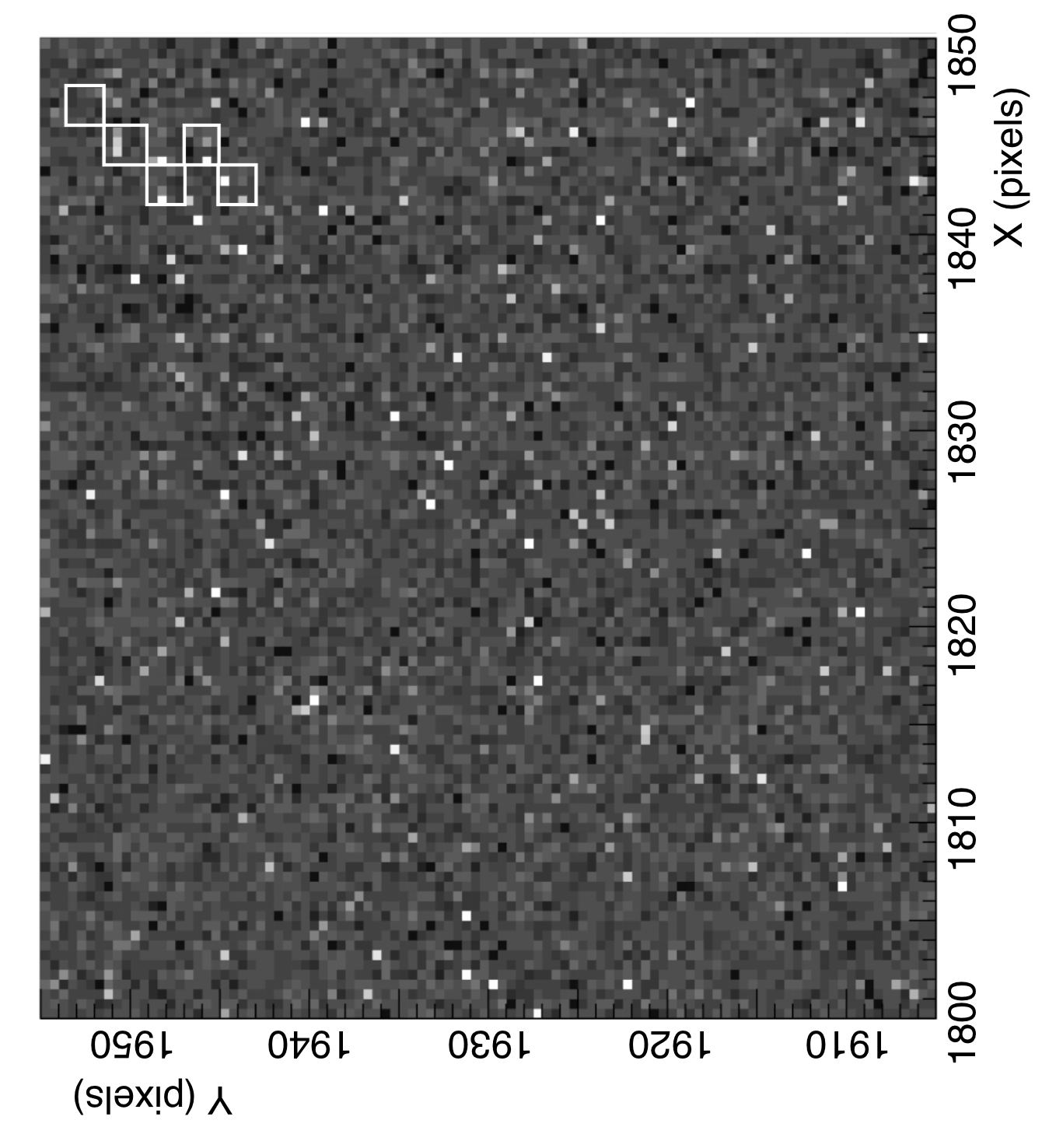}
\caption{Example of the {\it ghost-clusters} reconstructed with a 
total light larger than 200~photons in the {\it blind run}. 
White tiles represent the "macro-pixels" found to be over-threshold. Their geometrical distribution and 
the distribution of light in the original pixels
is quite different from the signal clusters shown in Fig.~\ref{fig:spot}.}
\label{fig:ghost-cluster}
\end{figure}

Figure~\ref{fig:ghost-cluster} shows an example of a {\it ghost-cluster} with a total light of 225 photons. As it can be easily seen also comparing to Fig. \ref{fig:spot}, some of the image characteristics (e.g. topological light distribution) can be further exploited to discriminate clusters induced by signals from {\it ghost-clusters}. This information has not been used in the data analysis presented in this paper, but it will provide additional handle to suppress the backgrounds in the future.

\subsection{Noise outside the sensitive volume}

\begin{figure}[htbp]
\centering
\includegraphics[width=.55\textwidth]{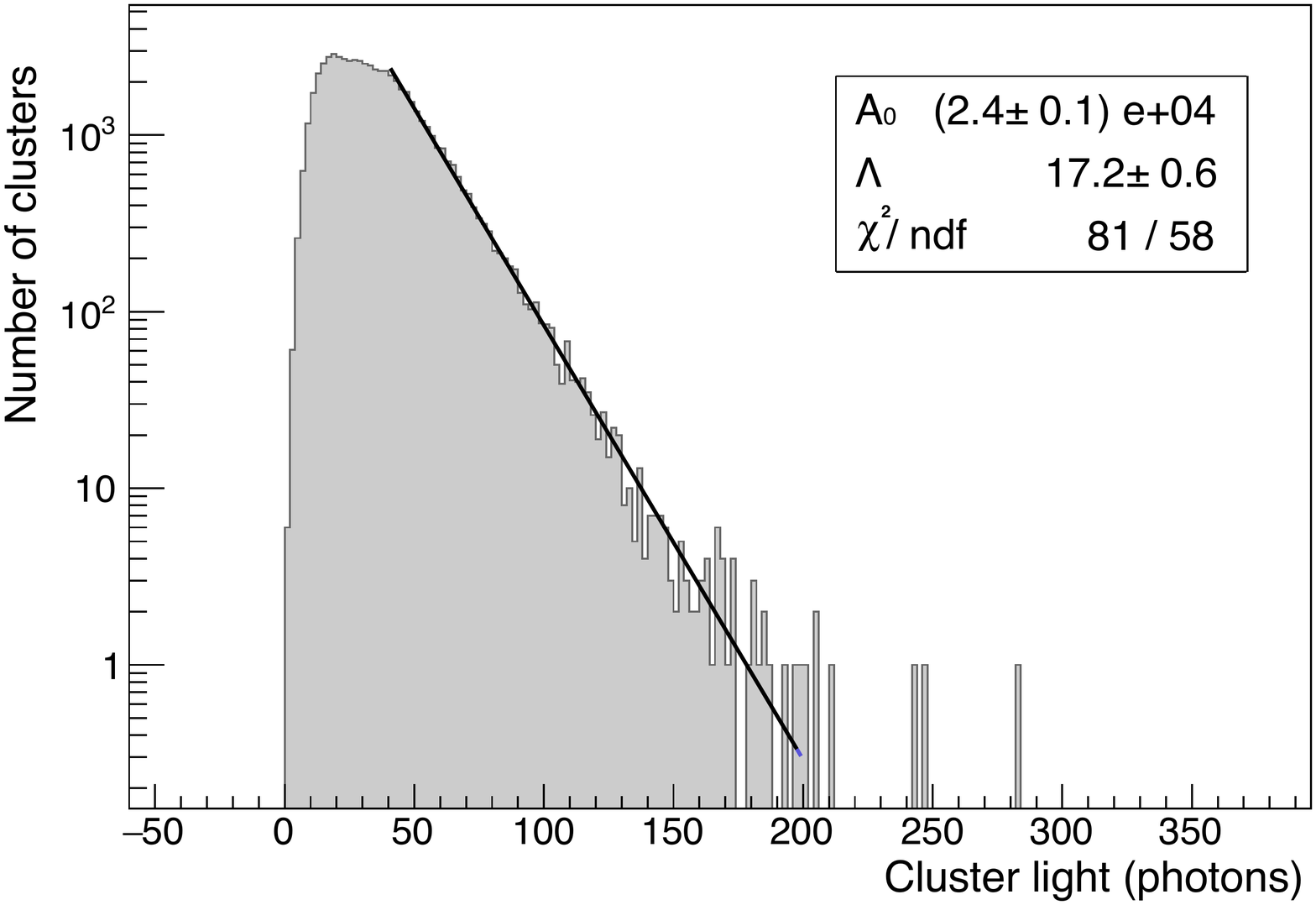}
\caption{Distribution of the light recorded clusters reconstructed outside from the sensitive area in a run with $^{55}$Fe source
with superimposed exponential fit (eq. \ref{eq:exp}). 
\label{fig:hq_out}}
\end{figure}

The GEM structure can, in principle, create a diffused light background because of possible micro-discharges. 
To evaluate it, the distribution of the light in the clusters reconstructed outside the sensitive area was studied. 
As it is shown in Fig.~\ref{fig:hq_out}, the obtained distribution is similar to the one due to the sensor electronic noise and has a tail that can be described with an exponential with a slope, $\Lambda=17.2~\pm~0.6$~photons, very similar to the one obtained for the sensor noise (see Fig.~\ref{fig:spot}).
The few events found outside the bulk of the distribution are short tracks very likely due to events occurred close to the GEM, where the residual electric field of the GEM is able to capture electrons and drive them toward the multiplication channels.

\subsection{Noise within the sensitive area}

\begin{figure}[htbp]
\centering
\includegraphics[width=.85\textwidth]{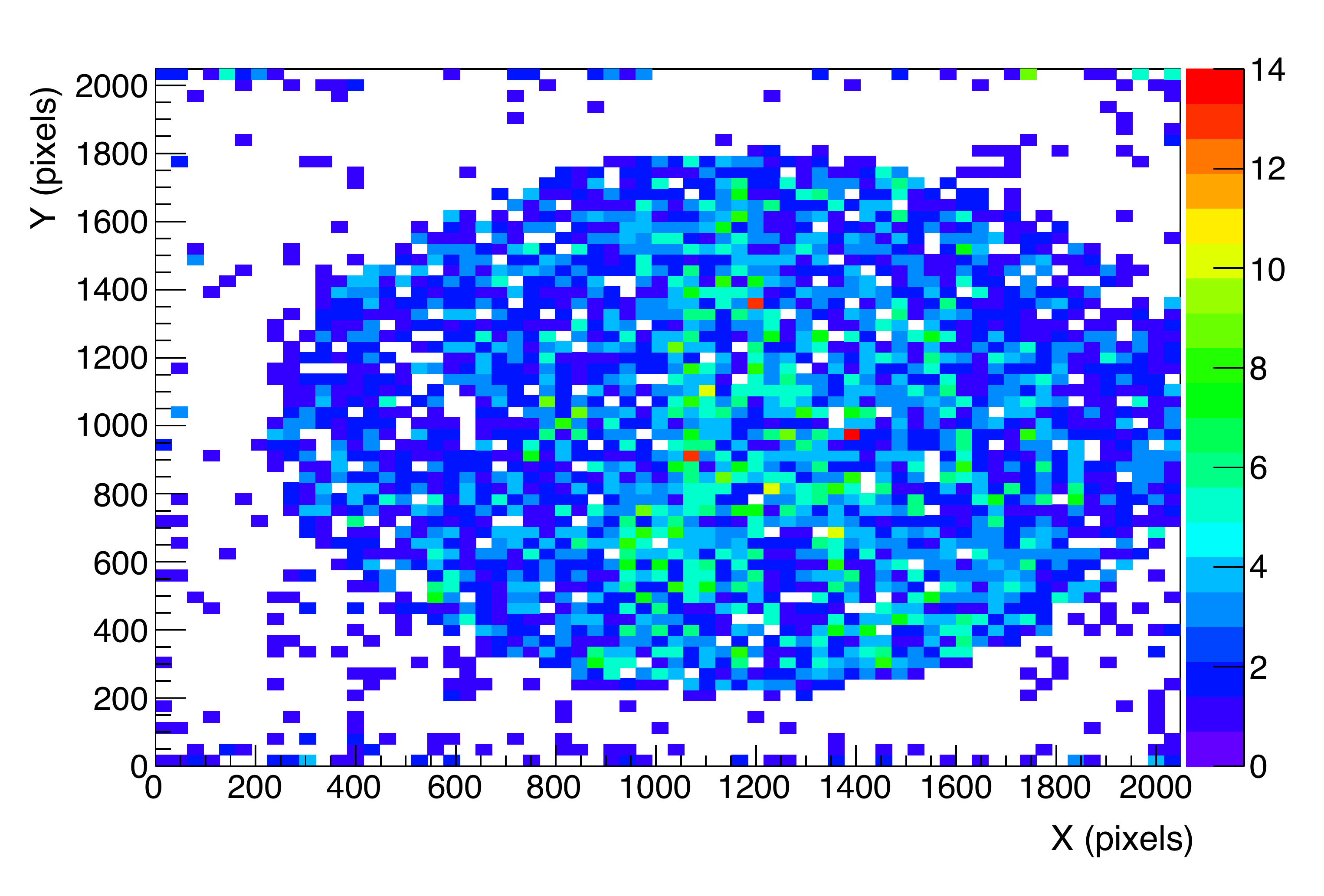}
\caption{Map of the positions of the reconstructed clusters for a run without the $^{55}$Fe source. \label{fig:map_bkg}}
\end{figure}

As already described in Sec.~\ref{sec:daq}, within the FC area an evident diffused and flat background is visible.
To study it, a run without the radioactive source was acquired. The map of all the reconstructed clusters is shown in Fig.~\ref{fig:map_bkg} (a different orientation of the camera, partially cut the sensitive area).
The observed spatial distribution of clusters is similar to the one found in presence of the source, while only few clusters are found outside the sensitive volume.

\begin{figure}[htbp]
\centering
\includegraphics[width=.85\textwidth]{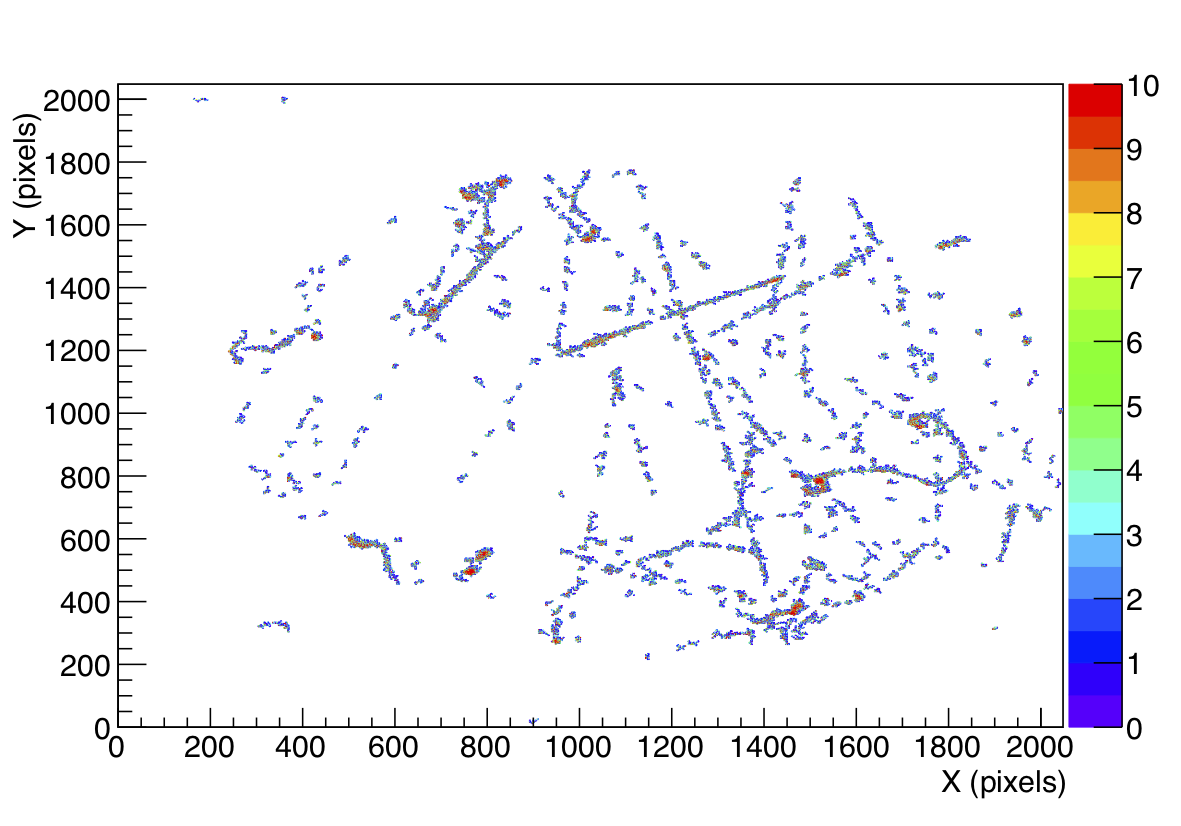}
\caption{Example of 10 events acquired in a run without the $^{55}$Fe source within the detectors. Color scale indicates the number of photons collected per pixel.}
\label{fig:bkg_evt}
\end{figure}

Figure~\ref{fig:bkg_evt} shows 10 overlapped events randomly chosen within the run.
They appear as to be mostly tracks due to cosmic rays
or low energy electrons from natural radioactivity.
In a radio-pure apparatus operating underground
such a background is expected to be strongly suppressed.
Moreover, pattern recognition should be able to identify and reject
residual events. For this reason, the effect of this background is not taken into account in this paper for the evaluation of the possible operative threshold.

\begin{figure}[htbp]
\centering
\includegraphics[width=.55\textwidth]{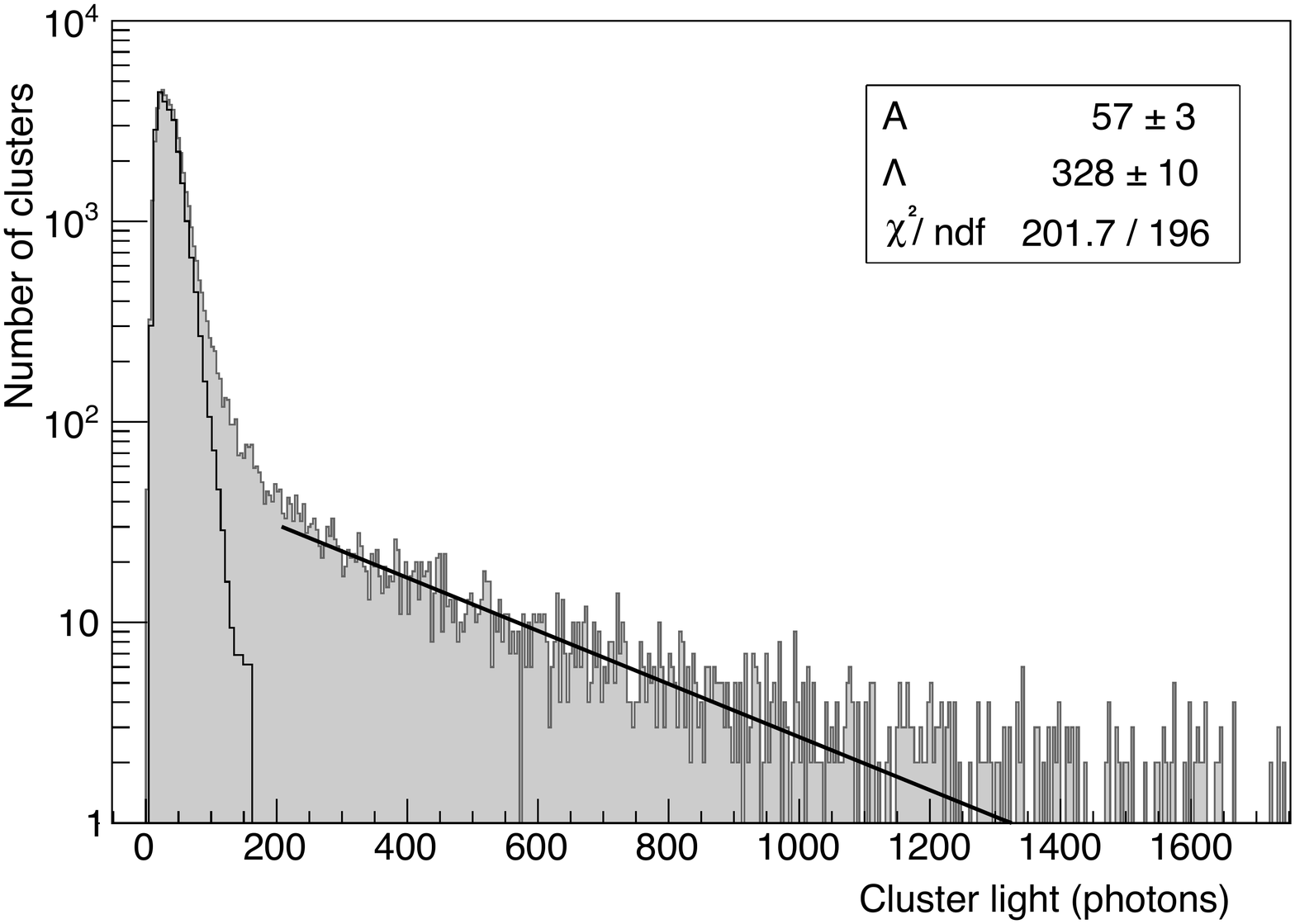}
\caption{Distribution of the light of background clusters with superimposed exponential fit (eq. \ref{eq:exp}). In black, the distribution of the light in {\it ghost clusters} is shown.}
\label{fig:hq_bkg}
\end{figure}

The distribution of the light of background clusters is shown in Fig.~\ref{fig:hq_bkg} with superimposed the distribution of light of {\it ghost-clusters}. The long tail on the right is therefore due to radioactivity events and is quite well described with an exponential function.

\section{Cluster size and light spectrum}

\begin{figure}[htbp]
\centering
\includegraphics[width=.27\textwidth, angle=270]{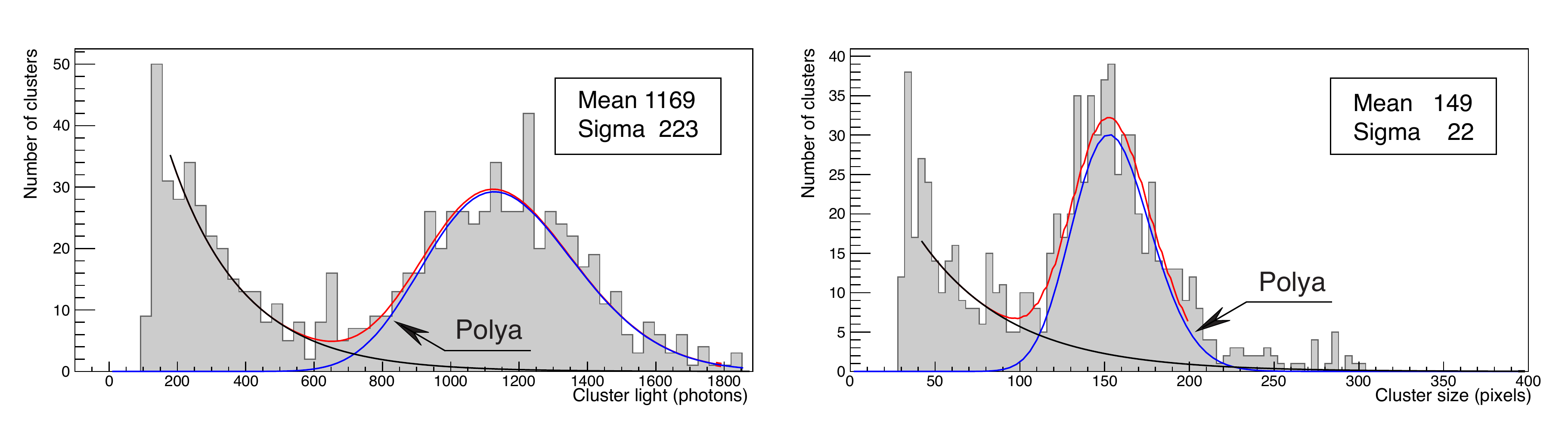}
\caption{Distribution of total light (left) and number of illuminated pixels (right) for a run taken with V$_{\rm GEM}$~=~450 V, E$_{\rm d}$~=~600 V/cm and E$_{\rm t}$~=~2 kV/cm.}
\label{fig:spectra}
\end{figure}

For each run, the spectrum of the total light in clusters reconstructed within the sensitive area and the distribution of their size (i.e. the number of over-threshold pixels) are studied.
Figure \ref{fig:spectra} shows an example of these distributions for a run taken with V$_{\rm GEM}$~=~450 V, E$_{\rm d}$ = 600~V/cm and E$_{\rm t}$ = 2~kV/cm.

In order not to slow-down the analysis procedure, only clusters with at least 30 pixels over-threshold were considered. The distribution is fitted with the sum of an exponential function to model the background due to natural radioactivity (Sec.~\ref{sec:bkg}), and a Polya function, expressed by Eq.~\ref{fun:polya}, often used to describe the response spectrum of MPGD \cite{bib:rolandiblum}:
\begin{equation}
   P(n)=\frac{1}{b\overline{n}}\frac{1}{k!}\left(\frac{n}{b\overline{n}}\right)^k \cdot e^{-n/b\overline{n}}
\label{fun:polya}
\end{equation}
%
where $b$ is a free parameter and $k=1/b-1$. The distribution has $\overline{n}$ as expected value, while the variance is governed by its mean and the parameter $b$: $\sigma^2=\overline{n}(1+b\overline{n})$. 
From the result of the fits it is possible to evaluate:
\begin{itemize}
\item the expected value of the distribution $\overline{n}$ and its variance $\sigma^2$. These parameters, when fitted on the light distributions, give the detector response in term of number of photons and the energy resolution. The latter will thus indicate the 
variance ($\sigma$) of the Polya fit in the whole paper. 
When fitting the number of illuminated pixels distribution, the average size of the clusters can be evaluated by taking into account the effective area of $130~\times 130~\mu$m$^2$ (see Sect. \ref{sec:daq}) acquired by each single pixel.
\item the integral of the Polya component, that is proportional to the total number of reconstructed clusters and that can be used to evaluate the detection efficiency;
\end{itemize}
Since, as it is shown on the left of Fig.~\ref{fig:spectra}, in this configuration 1169~$\pm$~223 photons are collected per cluster (i.e. each 5.9~keV released), a threshold of 400 photons corresponds to about 2~keV released in the sensitive volume.
The average cluster size was found to be 149 pixels (Fig.~\ref{fig:spectra}, right). 

\section{Results}
The response of the detector to the $^{55}$Fe source has been studied as a function of three different operative parameters: V$_{\rm GEM}$, E$_{\rm d}$ and E$_{\rm t}$. 
Main results are reported in this section.

\subsection{Dependence on the voltage applied to the GEM (V$_{\rm GEM}$)}

\begin{figure}[htbp]
\centering
\includegraphics[width=.42\textwidth]{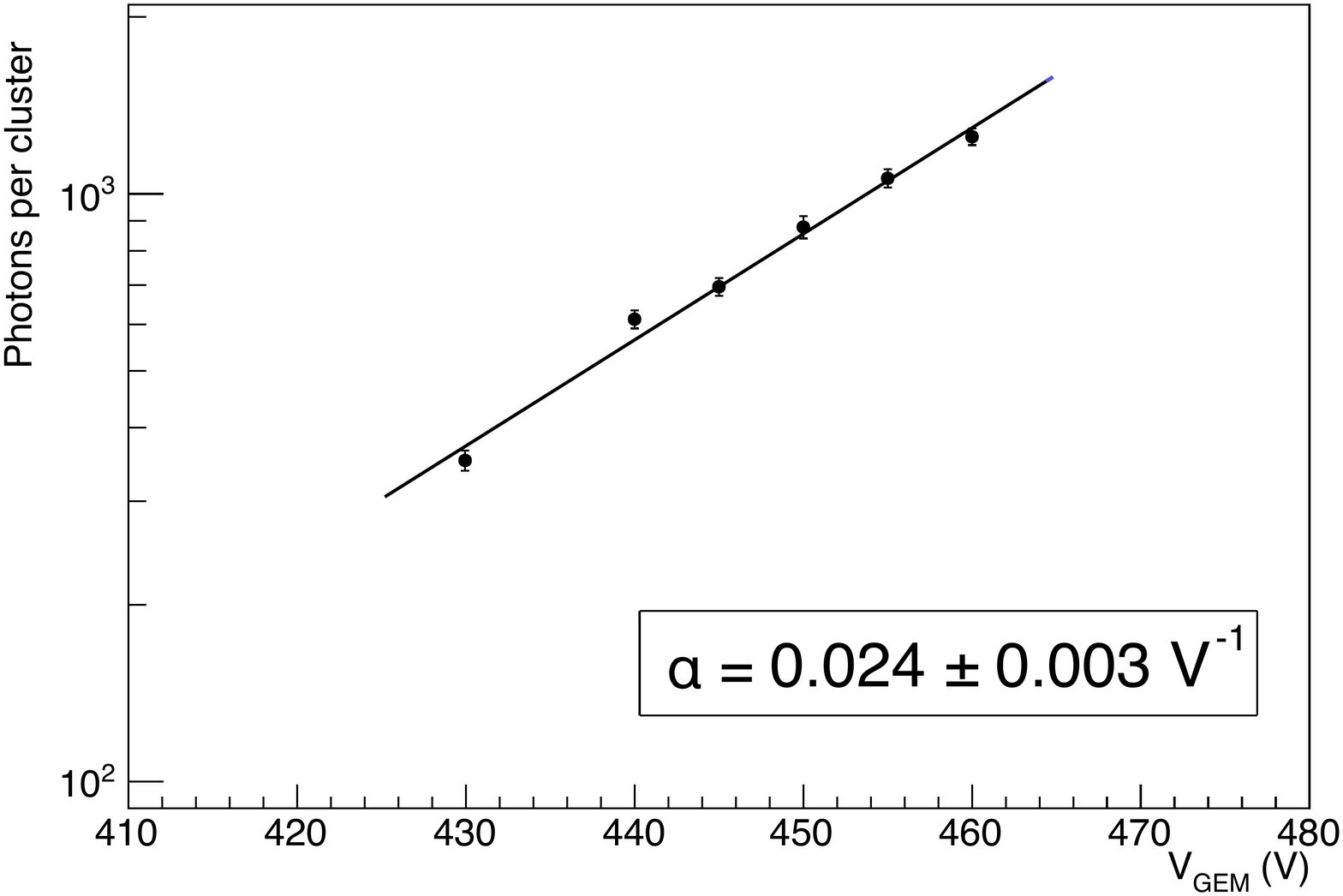}
\includegraphics[width=.435\textwidth]{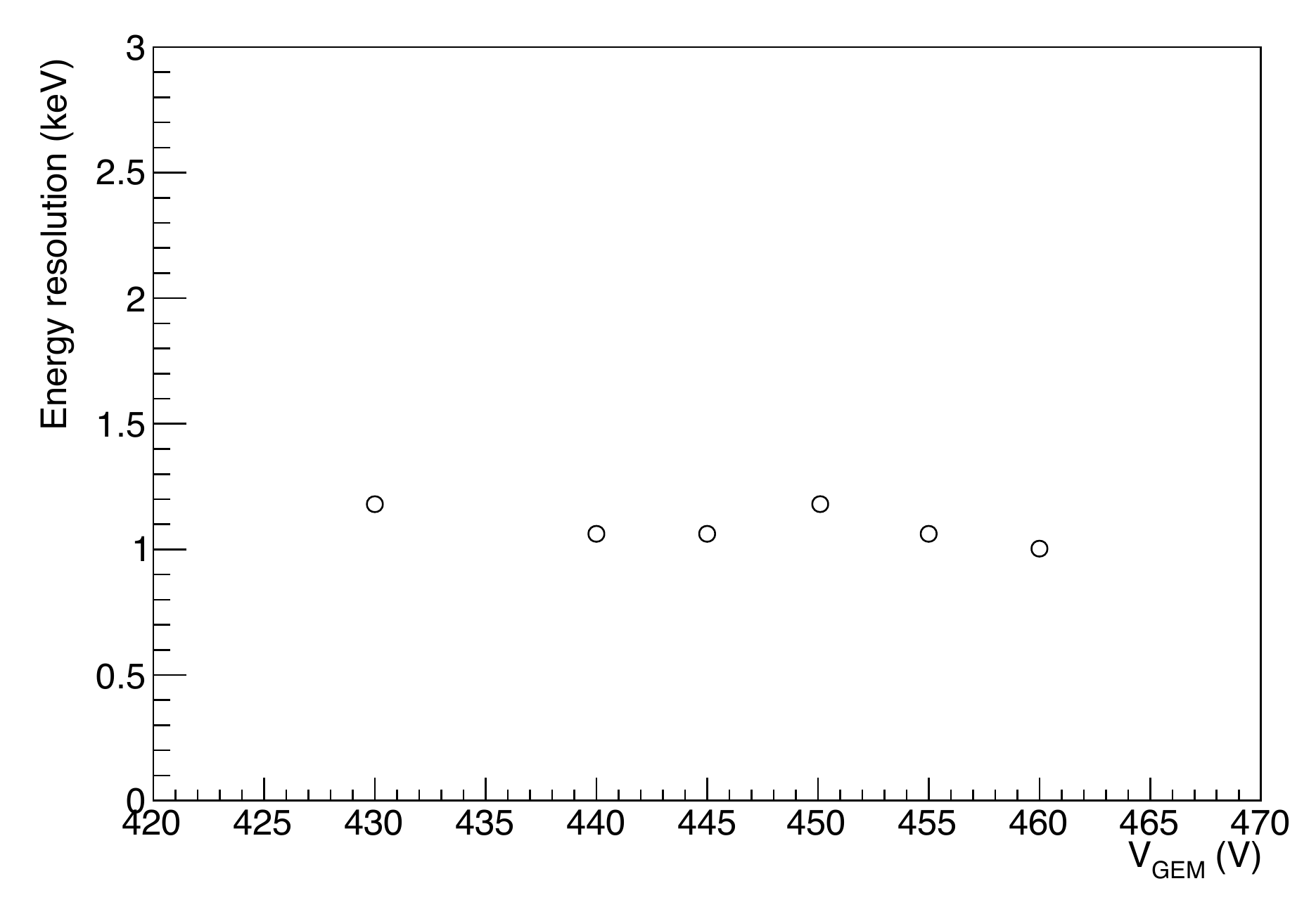}
\caption{Average of the light spectrum with an exponential fit (left) and its relative fluctuations (right) as a function of V$_{\rm GEM}$ for a run taken with E$_{\rm d}$ =~600 V/cm and E$_{\rm t}$ =~2 kV/cm.}
\label{fig:vgem1}
\end{figure}

The main parameter that determines the gain and therefore the light yield of the structure is the voltage applied to the GEM (V$_{\rm GEM}$).  
Figure \ref{fig:vgem1} shows the behavior of the average number of collected photons per cluster and its fluctuations as a function of V$_{\rm GEM}$.

From the superimposed fit to an exponential function:
\begin{equation}
    {\rm L} = {\rm C}e^{\alpha{V}}
\label{func:myExp}
\end{equation}

it is possible to evaluate that detector light yield increases exponentially and doubles every V$_{\rm GEM} =$~30 V step, showing 
the same behavior measured with a different prototype with 450 MeV electrons from beam \cite{bib:jinst_orange1}. The energy resolution is found not to be significantly dependent on the voltage applied to the GEM with a value around 20\% in good agreement with results obtained with similar experimental setup \cite{bib:loomba}. Even if these values are slightly higher than what can be obtained with an optimised
charge readout (see for example \cite{bib:ref6}), the big advantage of optical readout is the very high granularity allowing the reconstruction of cluster shapes, that can provide a crucial handle in signal identification and background rejection.

\begin{figure}[htbp]
\centering
\includegraphics[width=.45\textwidth]{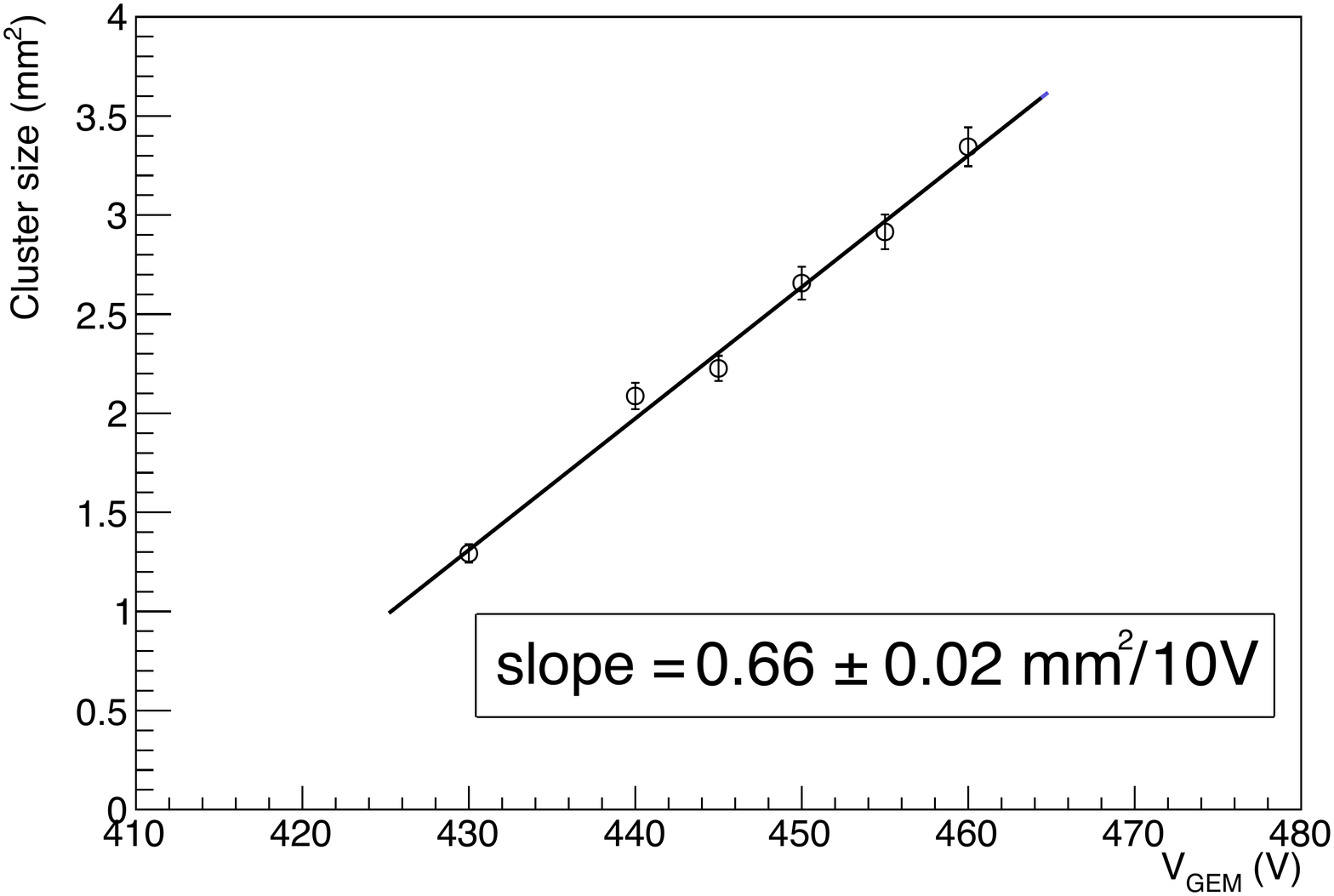}
\includegraphics[width=.45\textwidth]{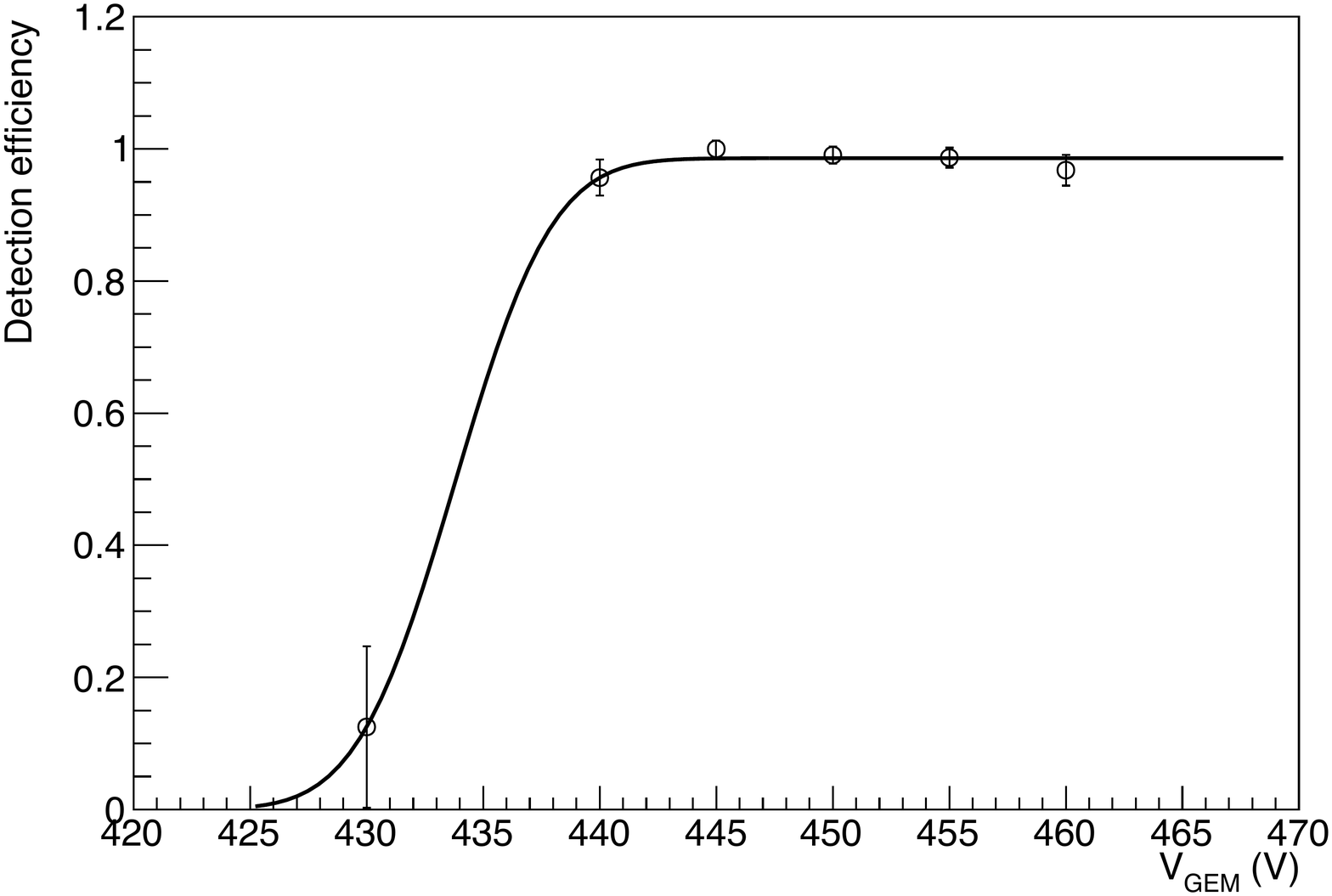}
\caption{Behavior of the dimension spectra (right) and detection efficiency as a function of V$_{\rm GEM}$ for a run taken with E$_{\rm d}$~=~600~V/cm and E$_{\rm t}$~=~2 kV/cm.}
\label{fig:vgem2}
\end{figure}

The cluster size increases with the GEM photon yield as it is shown on the left in Fig.~\ref{fig:vgem2}: a linear dependency with a slope of 0.66~mm$^2$/10~V is found. On the right of Fig.~\ref{fig:vgem2}, the total amount of the cluster detected normalised to its maximum value is shown.
Since it reaches a plateau and does not increase for V$_{\rm GEM}$ larger than 440~V, it is possible to conclude that this quantity represents also the global detection efficiency that is close to unity.

\subsection{Dependence on the Drift Field (E$_{\rm d}$)}

All measurements were taken with the $^{55}$Fe source 18 cm away from the readout plane and therefore, the response of the detector as a function of the electric field within the FC (E$_{\rm d}$) provides information on the effect of electron attachment and diffusion in our configuration.

\begin{figure}[htbp]
\centering
\includegraphics[width=.45\textwidth]{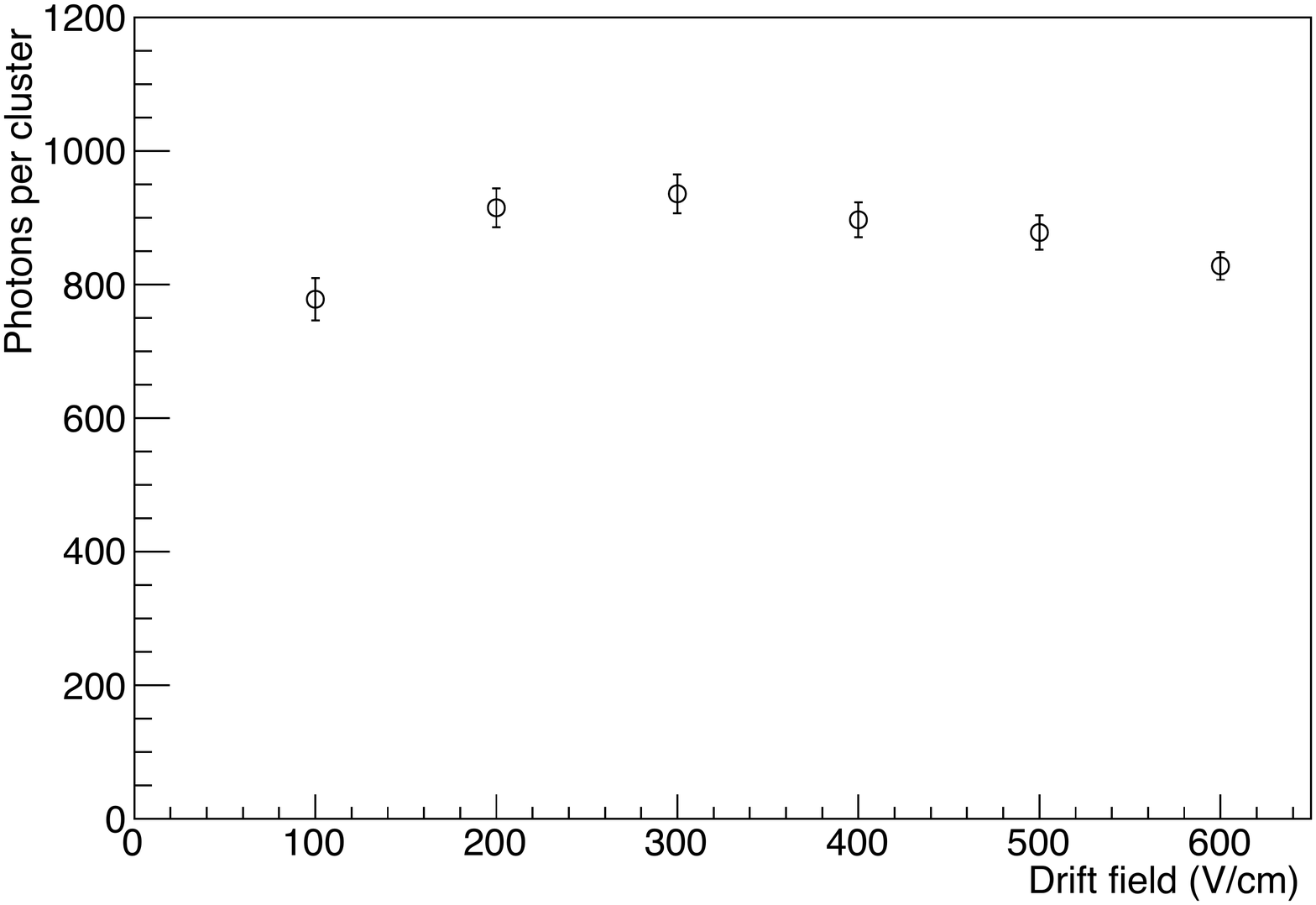}
\includegraphics[width=.45\textwidth]{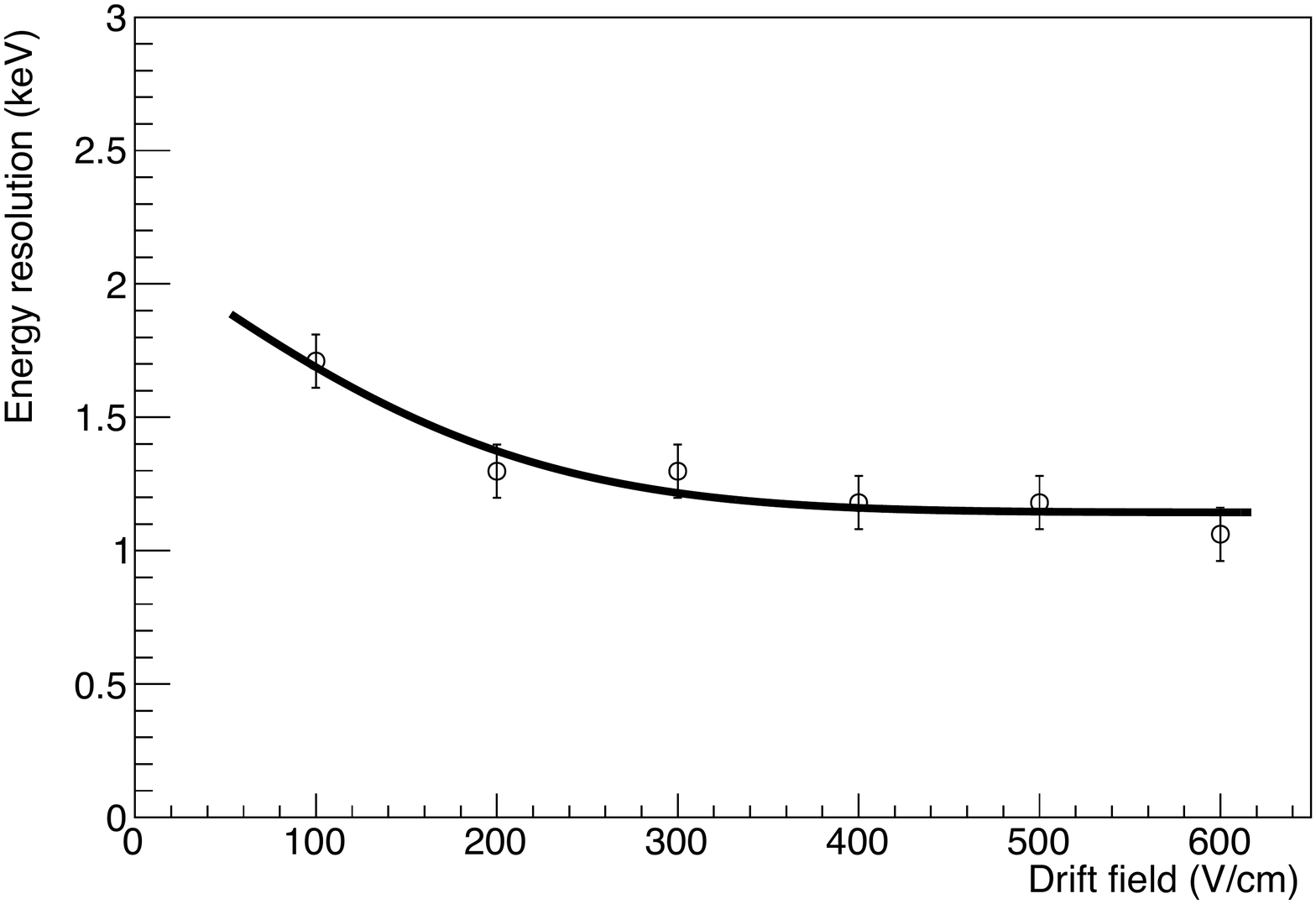}
\caption{Behavior of average light spectrum (left) and its relative fluctuations (right) as a function of E$_{\rm d}$ for a run taken with V$_{\rm GEM}$~=~450 V and E$_{\rm t}$ =~2 kV/cm.}
\label{fig:Ed1}
\end{figure}

The measured number of photons per reconstructed cluster (Fig.~\ref{fig:Ed1}, left) is quite stable.
A slight decrease is visible likely due to the de-focusing effect of the increasing drift field \cite{bib:thesis}.
Fluctuations of the number of photons per cluster (Fig. \ref{fig:Ed1}, right) have a small increase for small values of E$_{\rm d}$ (from around 20\% to almost 30\%).
\begin{figure}[htbp]
\centering
\includegraphics[width=.45\textwidth]{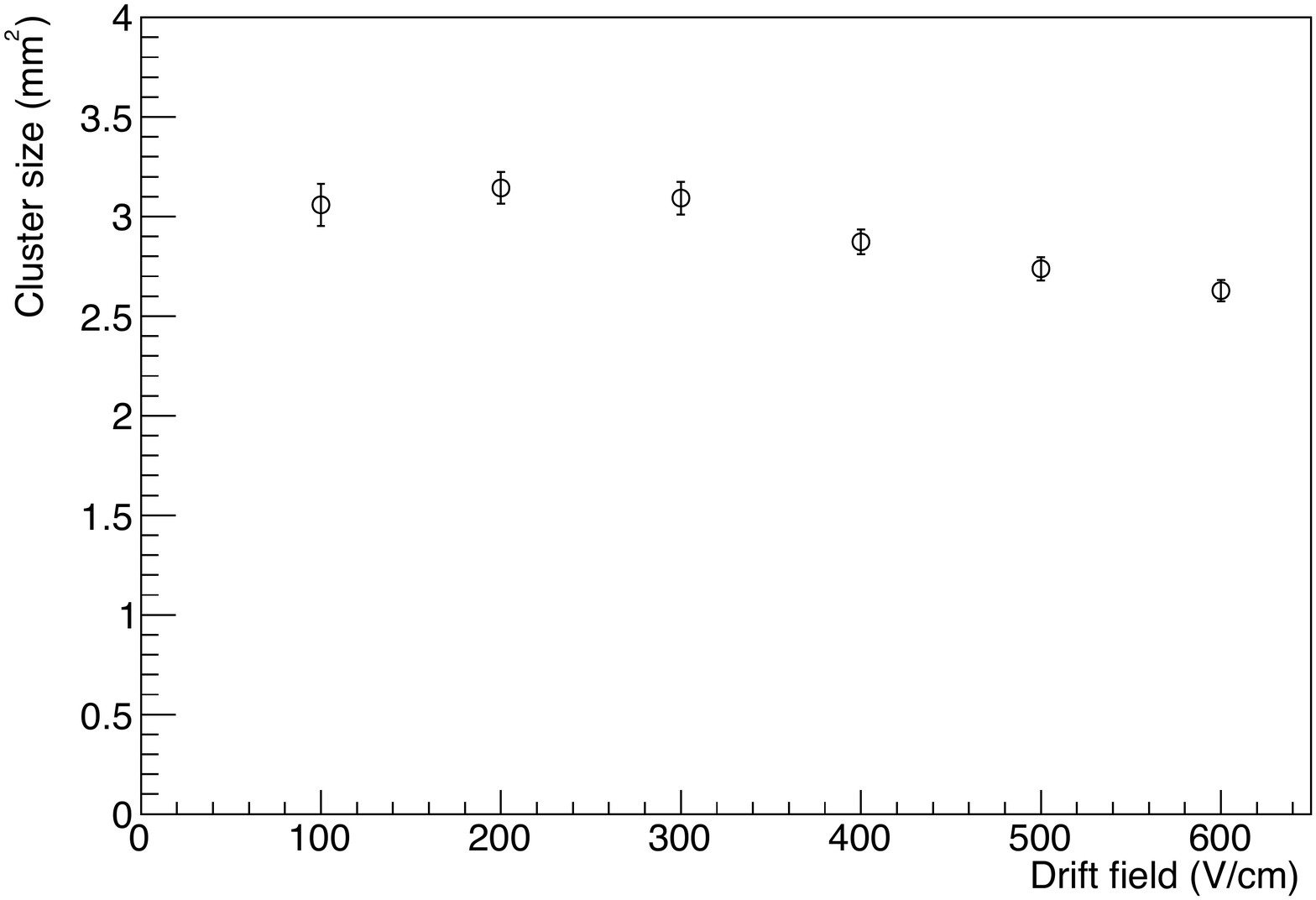}
\includegraphics[width=.45\textwidth]{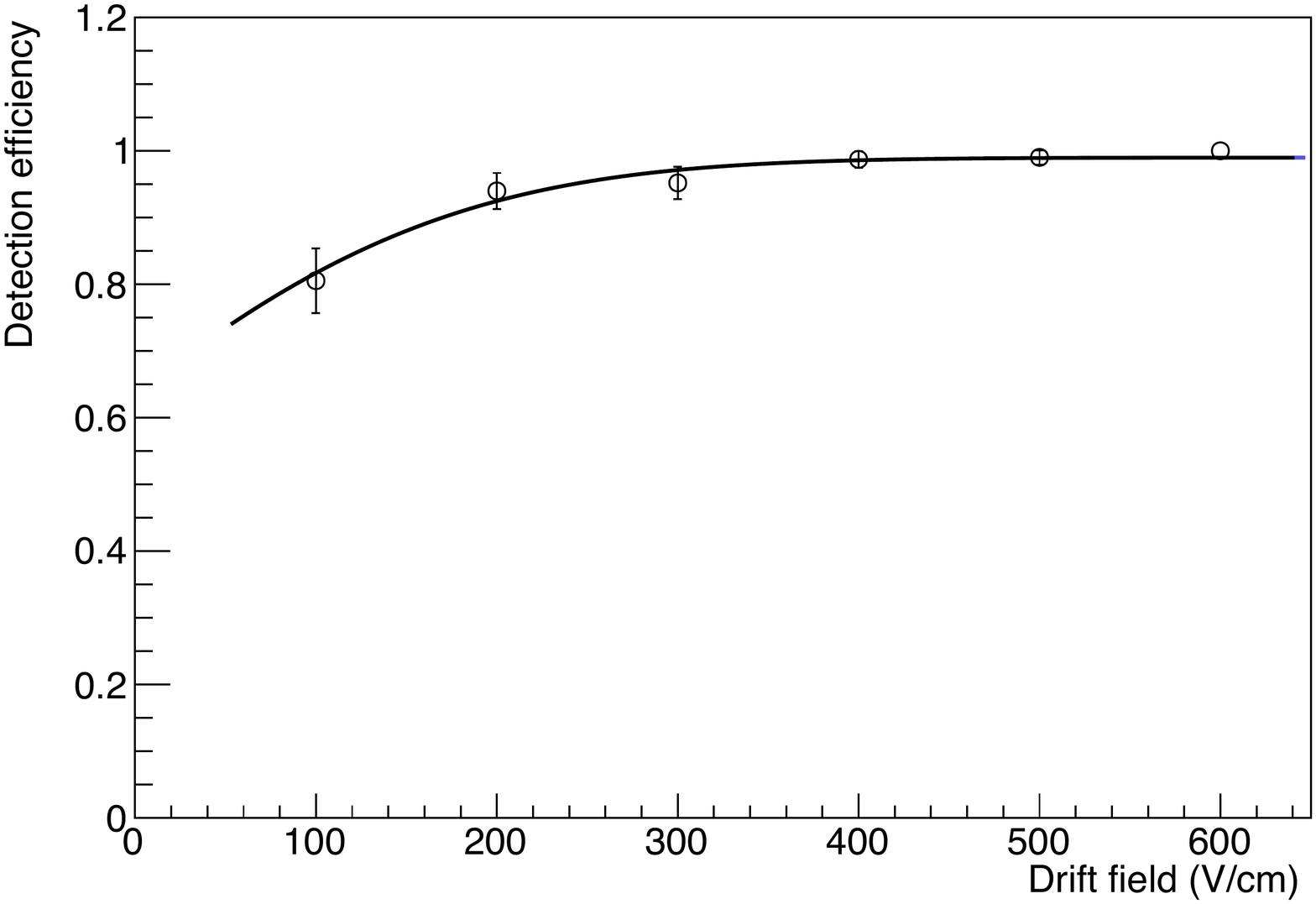}
\caption{Behavior of average cluster size (left) and detection efficiency (right) as a function of E$_{\rm d}$ for a run taken with V$_{\rm GEM}$~=~450 V and E$_{\rm t}$ =~2 kV/cm.}
\label{fig:Ed2}
\end{figure}
The size of detected clusters (left in Fig.~\ref{fig:Ed2}) decreases slowly as a function of E$_{\rm d}$ because of a smaller electron diffusion in gas. The detection efficiency (right of Fig.~\ref{fig:Ed2}) remains well above 95\% for E$_{\rm d}$ larger than 300 V/cm.

\subsection{Dependence on the Transfer Field (E$_{\rm t}$)}

The electric field in the gap between the GEM, E$_{\rm t}$, plays a crucial role in the electron transport and on the effective gain of the detector.

\begin{figure}[htbp]
\centering
\includegraphics[width=.45\textwidth]{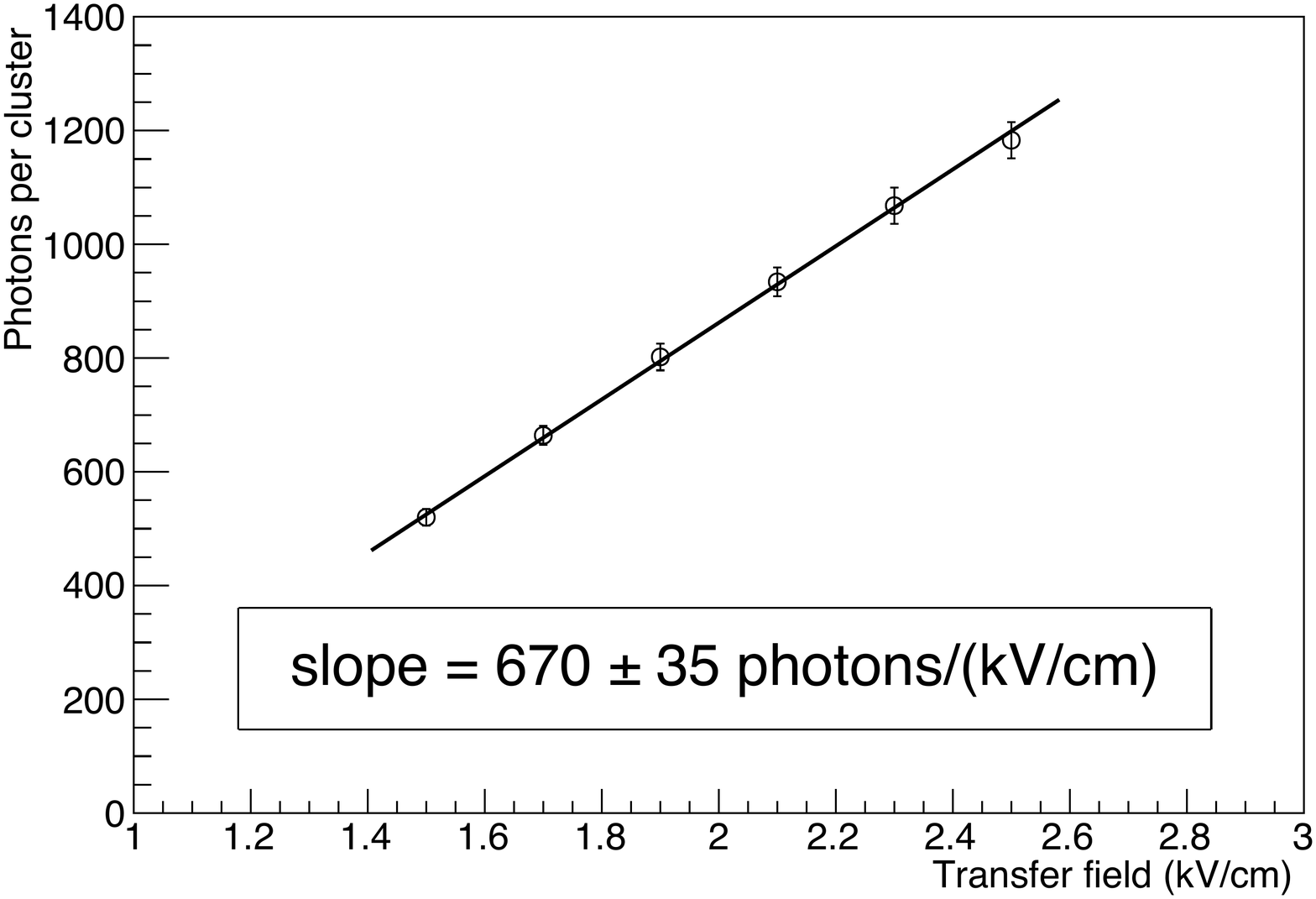}
\includegraphics[width=.45\textwidth]{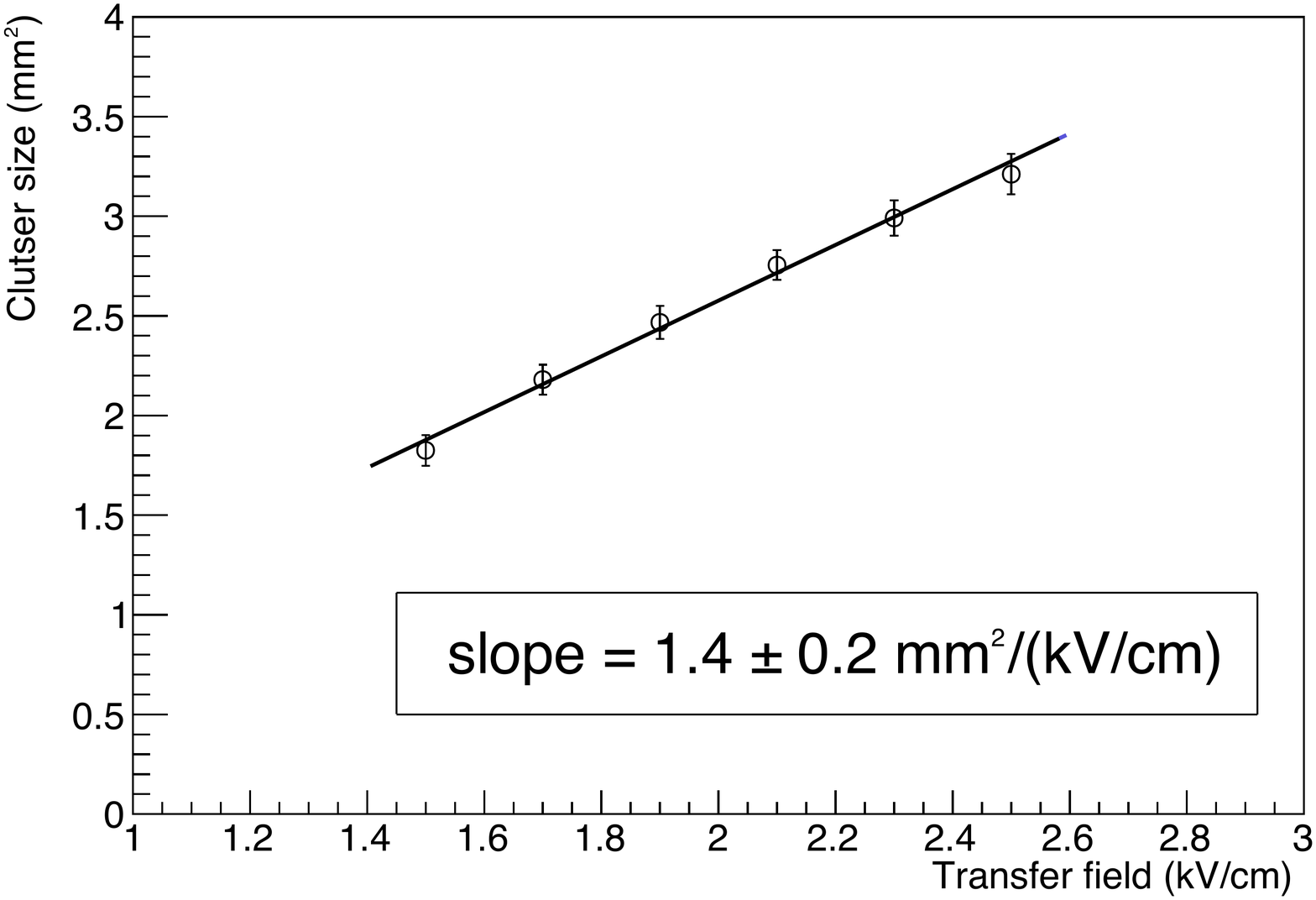}
\caption{Behavior of average light spectrum (left) and its relative fluctuations (right) as a function of E$_{\rm t}$ for a run taken with V$_{\rm GEM}$~=~450 V and E$_{\rm d}$~=~600~V/cm.}
\label{fig:Et1}
\end{figure}

Because of a better capability in extracting electrons \cite{bib:thesis}, the number of collected photons (Fig. \ref{fig:Et1}, left) increases linearly with the E$_{\rm t}$ reaching a value of 1200 for E$_{\rm t}$ = 2.5 kV/cm (while their fluctuations are quite stable around 20\%). 
In this configuration, therefore, a sensitivity of 0.2 collected photons per released keV was measured.

Also the size of detected clusters (right in Fig.~\ref{fig:Et1}) shows an almost linear increase as a function of E$_{\rm t}$ as a result of the increasing of the GEM light yield.

\section{Conclusion}

The analysis of the tests performed on the LEMON detector with the 5.9 keV photons
provided an important characterisation of the detector response. With a suitable field configuration (V$_{\rm GEM}$~=~460V, E$_{\rm d}$~=~600~V/cm and E$_{\rm t}$~=~2.5 kV/cm), the response of the detector is measured to be 1200 ph/cluster, i.e 1 photon each 5 released elettronvolts.
From the studies of the sensor intrinsic noise, it was possible to determine that a threshold of 400 photons ensures a rate of fake events lesser than 10 per year. With a sensitivity of 0.2 ph/eV this would represent a threshold of 2 keV.
With an E$_{\rm t}$~=~2 kV/cm, the detection efficiency was estimated to be well above 95\% down to V$_{\rm GEM}$~=~430 V where 1/3 of light is collected compared to V$_{\rm GEM}$~=~460 V and E$_{\rm t}$~=~2.5 kV/cm.
Therefore, working with the latter settings would provide 3 more times light and thus, full detection efficiency seems possible for 2 keV signals.

All these studies indicate that the described technique allows a full efficiency and very low sensor noise conditions with an operative threshold of 2 keV.

\bibliography{fe55}{}
\bibliographystyle{ieeetr}

\end{document}